\documentstyle[11pt,newpasp,twoside]{article}
\def\edcomment#1{\iffalse\marginpar{\raggedright\sl#1\/}\else\relax\fi}
\reversemarginpar

\begin{document}
\title{Magnetic Fields in Clusters of Galaxies Obtained Through the
Study of Faraday Rotation Measures}
\author{Federica Govoni}
\affil{Dip. Astronomia, Univ. Bologna, Via Ranzani 1, I-40127 Bologna, Italy}
\author{Luigina Feretti, Matteo Murgia}
\affil{IRA-CNR, via Gobetti 101, I-40129 Bologna, Italy}
\author{Greg Taylor}
\affil{NRAO, Socorro, NM 87801, USA}
\author{Gabriele Giovannini, Daniele Dallacasa}
\affil{Dip. Astronomia, Univ. Bologna, Via Ranzani 1, I-40127 Bologna, Italy}

\begin{abstract}
We are developing a new approach to investigate cluster 
magnetic field strengths and structures.
It is based on the comparison of 
simulated Rotation Measure and radio halo images,
obtained from 3-dimensional multi-scale cluster magnetic fields models, with 
observations. This approach is applied to A2255 in which wide diffuse radio
emissions and Rotation Measure images show the evidence of cluster magnetic 
fields on both large and small scales.
\end{abstract}

\section{Introduction}
The existence of cluster magnetic fields has been
demonstrated by different techniques.

Some clusters of galaxies show the presence
of wide diffuse radio emissions (radio halos and/or relics) associated
with the intra-cluster medium rather than a  
cluster galaxy.
Under minimum energy assumptions, it is possible to calculate
an equipartition magnetic field strength averaged over the 
entire halo volume. 
These estimates give $H_{eq}$$\simeq$0.1-1 $\mu$G (e.g. 
Feretti \& Giovannini 1996; Govoni et al. 2001a).

In a few cases, clusters with radio halos
show a hard X-ray excess. This emission is usually interpreted in terms of 
Inverse-Compton scattering of cosmic microwave background photons by the 
relativistic electrons responsible for the radio halo emission. In this case,
the measurements of the magnetic field 
strength (e.g. Fusco-Femiano et al. 1999, Rephaeli et al. 1999) 
inferred from the
ratio of the radio to X-ray luminosities are consistent with the 
equipartition estimates.

Indirect measurements of the cluster magnetic field 
strength can be determined in conjunction with the X-ray
observations of the hot gas, through the study     
of the Faraday Rotation Measure (RM) of radio
sources located inside or behind the cluster.
In clusters 
without cooling flows, 
RM studies of polarized radio sources   
lead to a magnetic field of 2-6 $\mu$G with
a correlation length in the range 2-15 kpc
(Feretti et al. 1995; Feretti et al. 1999;
Clarke et al. 2001;
Taylor et al. 2001; Govoni et al. 2001b).

The magnetic field strength obtained by RM studies is therefore 
higher than the value derived from the radio data 
and from Inverse-Compton X-ray emission in clusters with radio halos.
However, both methods are based on many simplifying assumptions 
(see e.g. a recent review by Carilli \& Taylor 2002).
Moreover, as pointed out by Newman et al. (2002), and Murgia
et al. (2002, in preparation) the observed RMs are usually interpreted in terms
of the simplest possible magnetic field model, and
previous estimates of the field by using RMs value are likely to 
be over-estimated. 

\section{The cluster A2255}
Here we present a preliminary analysis of the magnetic field
in the cluster A2255.

A2255 (z=0.0806; 1$''$$\simeq$2 kpc, with H$_0$=50 km/sMpc and q$_0$=0.5), 
is a suitable target for the investigation of magnetic fields
because it is characterized 
by the presence of well polarized radio galaxies
and contains both a radio halo and a relic source.

\begin{figure}
\plotfiddle{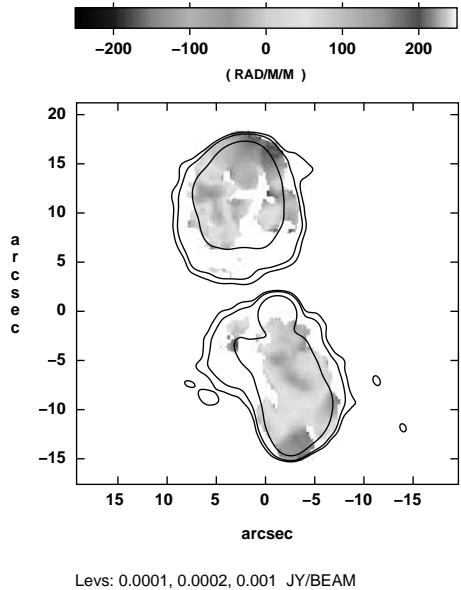}{7cm}{0}{35}{35}{-100}{-20}
\caption{RM image of the radio source $B1713+641$ in A2255,  
computed using the polarization angle images at
the frequencies 4535, 4885, 8085 and 8465 MHz with a resolution of 2$''$.
Contours refer to the total intensity image at 6 cm.
The values of RM range between $-100$~rad/m$^2$ and $210$~rad/m$^2$. 
The $<$RM$>$ is $67~$rad/m$^2$
and the $\sigma_{RM}$ is $59$~rad/m$^2$.
} 
\end{figure}

With high sensitivity 
and resolution ($\simeq$2$''$) 
Very Large Array (VLA) polarization data,
we obtained RM images of two extended 
radio galaxies belonging to the cluster. 
The RM images of these radio sources
indicates RM structures with
coherence lengths of about 10 kpc.
In Fig. 1 we report the RM image of one of these
sources.

The radio halo in A2255 was studied by several authors
(Jaffe \& Rudnick 1997, Harris et al. 1980, Burns et al. 1995, 
Feretti et al. 1997) with angular resolutions of tens of arcsec.
The extent of the radio emission implies the existence
of magnetic field on a much larger scale than implied
by the RM results.
Therefore,
the wide diffuse radio emissions and the Rotation Measure images 
in A2255 show the evidence of a cluster magnetic field 
both on large and small scales.
For this reason, to study the cluster magnetic fields it
is necessary to consider realistic magnetic field 
models where both small and large scale 
filamentary structures coexist.

\begin{figure}
\plotfiddle{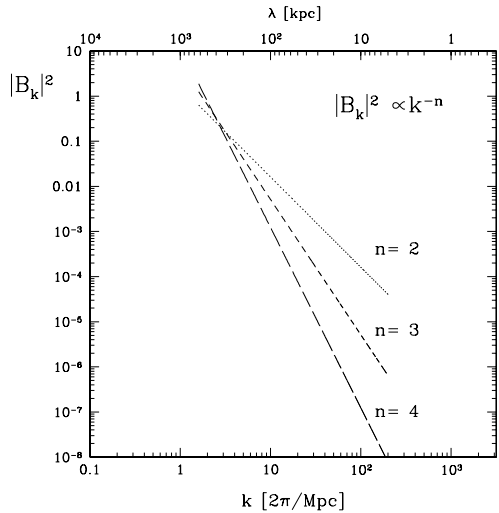}{6cm}{0}{35}{35}{-120}{-70}
\caption{
Power spectra of  
3-dimensional multi-scale magnetic field models used
in the simulations.
All the models are normalized to have the same magnetic field
energy. The magnetic field energy density decreases from 
the cluster center as the gas energy density, 
and the scale of the magnetic fields ranges from 5 kpc to 600 kpc.
} 
\end{figure}

\section{ Cluster Magnetic Field Models}

We developed a new approach to investigate
cluster magnetic fields by comparing simulated RM and radio halo images,
with observations.
We applied this method to A2255.
The FARADAY tool (Murgia et al. 2002,
in preparation) permits us to simulate RM and radio halo images
from 3-dimensional multi-scale cluster magnetic fields.
The magnetic field power spectrum of our
models follow the equation:
$|B_k|^2\propto k^{-n}$.
Different models will generate different magnetic field configurations 
and therefore will give rise to very different simulated RM and radio halo
images.

Fig.~2 shows 
power spectra of magnetic field models
used in the simulations.
All the models are normalized to have the same total magnetic field
energy. The magnetic field energy density decreases from 
the cluster center as the gas energy density, 
and the scale of the magnetic fields ranges from 5 kpc to 600 kpc.
The aim of this new approach is to find the optimal
magnetic field strength
and structure capable of describing both the small scale structures
(of about 10 kpc) seen in the RM images, 
and the large scale features visible in the radio halo of A2255
(Feretti et al. 1997, see Fig. 3, right on the top).

\begin{figure}
\plotfiddle{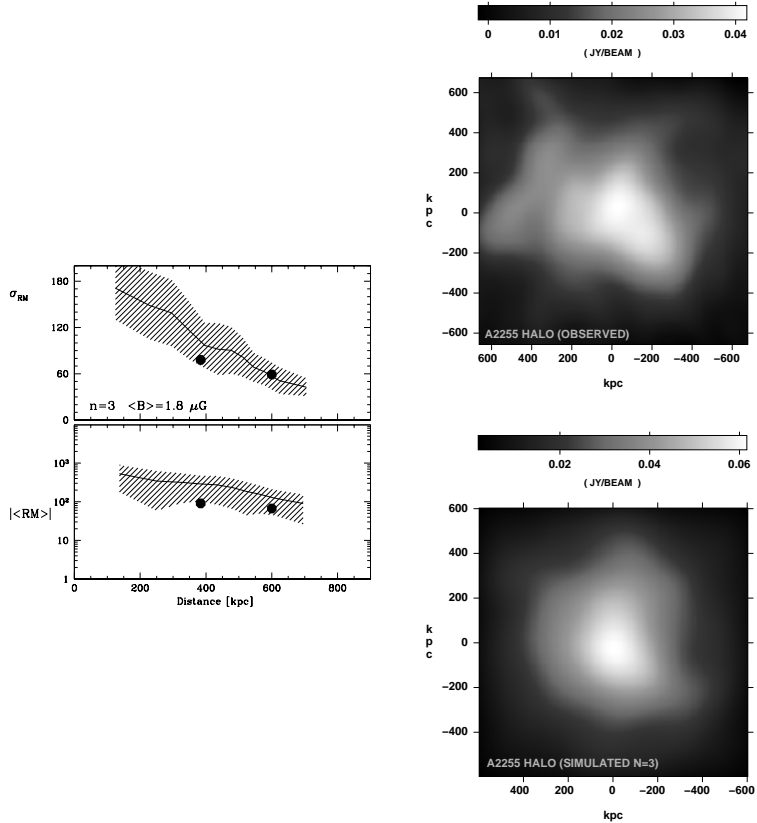}{10cm}{0}{50}{50}{-140}{-110}
\caption{
Simulation results
obtained with a power spectrum index $n=3$.
Left: RM data points ($\sigma_{RM}$ and $|<$RM$>|$)
obtained in A2255 (black points) 
are compared with the simulation (shaded region).
Right Top: WSRT radio image at 90 cm with resolution 
of $\simeq 170$ kpc
of the halo source in A2255 obtained by Feretti et al. (1997).
Right Bottom: Simulated large scale radio features 
as expected at the same frequency and resolution 
of the WSRT image.
} 
\end{figure}

The best power spectrum index which can explain 
both the RM data and the radio halo structure seems $n=3$.
Fig. 3 compares the results from a simulation 
with a power spectrum index $n=3$ with the observations.
On the left the RM data points ($\sigma_{RM}$ and $|<$RM$>|$)
obtained in A2255  
are compared with the expectation of the simulations (shaded region).
On the right the Westerbork Synthesis Radio Telescope
(WSRT) radio image at 90 cm with a linear resolution of about 170 kpc
(Feretti et al. 1997)
is compared with the simulated large scale radio halo 
as expected at the same frequency and resolution.
This steep magnetic field power spectral index is consistent 
with that (i.e. $n$=2.7) found by Dolag et al. (2002)
in cosmological magnetic hydrodinamical 
simulations.

Simulations with power spectrum index $n=2$
explain the RM data, but
predict a much smoother than observed structure of the radio
halo,
while $n=4$ seems to be
ruled out since the $|<$RM$>|$ is overestimated by a large factor.
Moreover, since most of the magnetic field energy is in the
large scales, the model with $n=4$ 
fails to reproduce the small features seen in the RM images.

\section{Conclusions}
Observable radio data, compared with simulated RM and radio halo images
obtained from 3-dimensional multi-scale cluster magnetic field models,
can significantly improve the knowledge of the cluster
magnetic field structure and strength.  
This approach is applied to A2255 in which wide diffuse radio
emissions and Rotation Measure images show the evidence of cluster magnetic 
fields on both large and small scales.
We find a magnetic field power spectral index of 3 from scales of
5 to 600 kpc can reproduce both the RM structures and the 
large scale radio halo.

\end{document}